\documentstyle[12pt,epsfig]{article}                                                             
                                                             
\newlength{\dinwidth}                                                             
\newlength{\dinmargin}                                                             
\def\lapproxeq{\lower .7ex\hbox{$\;\stackrel{\textstyle                                                             
<}{\sim}\;$}}                                                             
\def\gapproxeq{\lower .7ex\hbox{$\;\stackrel{\textstyle                                                             
>}{\sim}\;$}}                                                             
\def\be{\begin{equation}}                                                             
\def\ee{\end{equation}}                                                             
\def\bea{\begin{eqnarray}}                                                             
\def\eea{\end{eqnarray}}                                                             
                                                             
\def\funp{{I\!\!P}} 
\newcommand{\porpbar} 
{\!\,^{\scriptscriptstyle(}$\mbox{$\bar{p}$}$\,^{\scriptscriptstyle)}}

\begin{document}                                                             
\titlepage                                                            
\begin{flushright}                                                             
IPPP/01/23 \\      
DCPT/01/46 \\                                                             
22 May 2001 \\                                                             
\end{flushright}                                                             
                                                            
\vspace*{2cm}                                                             
              
\renewcommand{\thefootnote}{\fnsymbol{footnote}}                                           
\begin{center}                                                             
{\Large \bf RAPIDITY GAP HIGGS SIGNAL AT THE TEVATRON AND THE LHC\footnote[2]{Presented 
at the XVth Workshop Les Rencontres de Physique de la Vall\'{e}e d'Aoste, La Thuile, 
Aosta Valley, 4-10 March, 2001.}} \\   
                                                            
\vspace*{1cm}                                                             
V.A. Khoze$^a$ \\                                                             
                                                           
\vspace*{0.5cm}                                                             
$^a$ Department of Physics and Institute for Particle Physics Phenomenology, University of      
Durham, Durham, DH1 3LE \\                          
\end{center}                                                            
                                                            
\vspace*{2cm}                                                             
                                                            
\begin{abstract}                                                             
We quantify the rate and the signal-to-background ratio for Higgs $\rightarrow b\bar{b}$ 
detection in double-diffractive events at the Tevatron and the LHC.  The signal is 
predicted to be very small at the Tevatron, but observable at the LHC.  We show that the 
double-diffractive dijet production may serve as a unique gluon factory.  This process 
can be used also as a Pomeron-Pomeron luminosity monitor.
\end{abstract}                                                  
             
\newpage                   
\section{Introduction}           

\renewcommand{\thefootnote}{\arabic{footnote}}
One of the biggest challenges facing the high-energy experiments is to find a good 
signal with which to identify the Higgs boson.  Following the closure of LEP2, the focus 
of searches for the Higgs is concentrated on the measurements at the present and 
forthcoming hadron colliders, the Tevatron and the LHC.

To ascertain whether a Higgs signal can be seen, it is crucial to show first that 
the background does not overwhelm the signal.  For instance, as well known, an 
observation of the inclusive intermediate mass Higgs production, that is $pp$ or 
$p\bar{p} \rightarrow HX$ with $H \rightarrow b\bar{b}$ is considered to be impossible 
because of an extremely small signal-to-background ratio due to gluon-gluon fusion, 
$gg \rightarrow b\bar{b}$.  One possibility which is widely discussed \cite{Spira} is to 
observe the Higgs in association with massive particles ($W/Z$, $t$-quarks).  Another 
way to reduce the background, which at first sight looks quite attractive, is to study 
the central production of the Higgs in events with a large rapidity gap on either 
side, see, for example, \cite{DKT}--\cite{KMR3}.  An obvious advantage of the 
rapidity gap approach is the spectacularly clean experimental signatures:  hadron-free 
(\lq no-flight\rq) zones between the remnants of the incoming protons and the produced 
system.

Let us recall \cite{DKT}--\cite{Bjorken} that, in hard production processes, a 
gap corresponds to a rapidity region devoid of QCD radiation and represents 
a footprint of the colour-singlet $t$-channel exchange (that is a Pomeron or $W/Z$ or photon).

Events with large rapidity gaps may be selected either by using a calorimeter or by 
detecting leading protons with beam momentum fractions $x_p$ close to 1.  If the momenta 
of the leading protons can be measured with very high precision then a centrally 
produced state may be observed as a peak in the spectrum of the missing-mass ($M$) 
distribution.  Indeed, it has recently been proposed \cite{AR} to supplement CDF with 
very forward detectors to measure both the proton and antiproton in Run II of the Tevatron 
in events with the fractional momentum loss $\xi = 1 - x_p \lapproxeq 0.1$ with extremely 
good accuracy, corresponding to missing-mass resolution $\Delta M \simeq 250$~MeV.  
The experimental proposal is focused on searches for the Higgs boson, on possible 
manifestations of the physics beyond the Standard Model, as well as on unique studies 
of some subtle aspects of QCD dynamics.  It is expected that, in association with the 
high $x_p$ protons and antiprotons, a central system with mass up to about 200~GeV can 
be produced at the Tevatron.

Turning to the LHC, the physics menu is extremely rich, and the studies of the central 
production processes, including Higgs boson searches, have some important advantages.  
Here we specialize on central $H \rightarrow b\bar{b}$ production\footnote{The 
discussion of $H \rightarrow WW^*/ZZ^*$ central production as a viable way to 
identify the intermediate mass Higgs can be found in Refs.~\cite{KPRZ,KMR4}.}.

In Section~2 we briefly recall the QCD mechanism for the double-diffractive production 
of a system of large invariant mass $M$.  We use this formalism in Section~3 to study 
the background for double-diffractive $H \rightarrow b\bar{b}$ production.  Then in 
Section~4 we present estimates for the rates of Higgs events with rapidity gaps at the 
Tevatron and the LHC.  Finally, in Section~5, we present our conclusions.

\section{Double-diffractive hard production processes}

Here we present the estimates of the cross-sections for high energy processes of the 
type
\begin{equation}
\label{eq:a1}
pp \; \rightarrow \; p \: + \: M \: + \: p,
\end{equation}
and similarly for $p\bar{p}$, where a \lq plus\rq\ sign indicates the presence of a 
large rapidity gap.  To be precise, we calculate the rate for the double-diffractive 
exclusive production of a system of large invariant mass $M$, say a Higgs boson.  
Our discussion below will be focused on the case of an intermediate mass Higgs which 
dominantly decays into the $b\bar{b}$ final state.  

From the outset we would like to make it clear that at present there is no consensus 
within the community regarding the evaluation of the double-diffractive production 
cross sections, see \cite{AR,KMR5}.  The literature shows a wide range of predictions 
varying by many orders of magnitude.  This is not so surprising, keeping in mind that 
the evaluation of the rate of double-diffractive hard production is in some sense analogous 
to an attempt to estimate the chances for two camels to go through the eye of a needle.  
In the latter \lq {\it biblical}\rq\ case, the outcome crucially depends on the size of the 
eye, the camel's height and on its elasticity.  In order to navigate in our ({\it back-to-earth}) 
diffractive world it is necessary to invoke a dynamical model for the Pomeron.

\begin{figure}[htb]
 \vspace{6.5cm}
\includegraphics{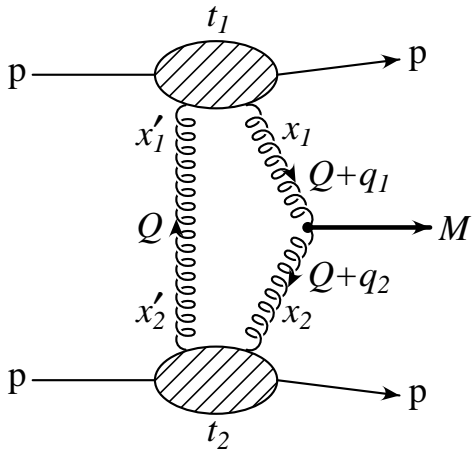}
 \caption{\it
 Schematic diagram of double-diffractive production of a system of 
 invariant mass $M$, that is the process $pp \rightarrow p + M + p$.
    \label{exfig} }
\end{figure}

One extreme possibility is the non-perturbative approach of Refs.~\cite{Schafer,BL} 
which exemplifies \lq\lq soft\rq\rq\ Pomerons (elastic camels).  Another extreme 
\cite{MS,LM} is to consider the so-called \lq\lq hard\rq\rq\ Pomeron (inelastic camel).  
The \lq\lq soft\rq\rq\ Pomeron-like models give
\begin{equation}
\label{eq:a2}
\frac{\sigma_{\rm max} (\funp \funp \rightarrow H)}{\sigma_{\rm incl} (gg \rightarrow H)} 
\; \sim \; (\sigma_{\rm el}/\sigma_{\rm tot})^2,
\end{equation}
where the suppression factor containing the elastic and total $pp$ cross  
sections is the probability of having two rapidity gaps, on either side of the Higgs.  
The low extreme, based on the \lq\lq hard\rq\rq\ Pomeron is
\begin{equation}
\label{eq:a3}
\frac{\sigma_{\rm min} (\funp\funp \rightarrow H)}{\sigma_{\rm incl} (gg \rightarrow H)} 
\; \sim \; (M_H^2 \sigma_{\rm tot})^{-2},
\end{equation}
where now the suppression factor is the probability to have a point-like  
dipole configuration (with the size of the eye $\lambda \sim 1/M_H$) in each Pomeron so 
that they have sufficient chance to fuse into the Higgs.  These simple estimates of the 
suppression factor range from $10^{-1}$ to $10^{-12}$.  Although naive, these results 
are, in fact, quite representative of the range of values that may be found in the 
literature.

As in Refs.~\cite{KMR1,KMR2,KMR3,KMR5,MRK} we adopt here the perturbative two-gluon 
exchange picture of the Pomeron, where the amplitude for the double-diffractive process 
is shown in Fig.~1.  The hard subprocess $gg \rightarrow M$ is initiated by gluon-gluon 
fusion and an additional relatively soft $t$-channel gluon is needed to screen the colour 
flow across the rapidity gap intervals.

A fundamental difference between the various theoretical approaches concerns the 
speficiation of the exchanged gluons.  {\it Either} non-perturbative gluons are used in which 
the propagator is modified so as to reproduce the total cross section \cite{BL,EML}, or a 
perturbative QCD estimate is made \cite{KMR2} using an unintegrated, skewed gluon density that 
is determined from the conventional gluon obtained in global parton analyses.  Note that the 
non-perturbative normalisation based on the value of the elastic or total cross section fixes 
the diagonal gluon density at $\hat{x} \sim \ell_T/\sqrt{s}$ where the transverse momentum 
$\ell_T$ is small, namely $\ell_T < 1$~GeV \cite{KMR1,KMR2,Berera}.  Thus the value of $\hat{x}$ 
is even smaller than
\begin{equation}
\label{eq:a4}
x^\prime \; \approx \; \frac{Q_T}{\sqrt{s}} \; \ll \; x \; \approx \; \frac{M}{\sqrt{s}},
\end{equation}
where the variables are defined in Fig.~1.  However, the gluon density grows as $x \rightarrow 
0$ and so the use of a non-perturbative normalisation will lead to an overestimation of 
double-diffractive cross sections.

It is important to emphasize that the rapidity gap signature is very promising but, at the 
same time, quite a fragile tool.  The gaps may easily fade away (filled by hadronic 
secondaries) due to various sources of QCD \lq\lq radiation damage\rq\rq:
\begin{itemize}
\item[(i)] soft rescattering of spectator partons caused by the transverse overlap of the 
two incoming protons (classic minimum-bias physics);
\item[(ii)] bremsstrahlung induced by the \lq active\rq\ partons in the hard subprocesses;
\item[(iii)] radiation originating from the small transverse distances in two-gluon Pomeron 
dipole.
\end{itemize}

A crucial numerical difference between the approaches concerns the size of the factor $W$ 
which determines the probability for the gaps to survive in the (hostile) QCD 
environment\footnote{We call $W$ the survival probability of rapidity gaps following Bjorken 
\cite{Bjorken}, who first introduced such a factor in the context of soft rescattering 
effects.}.

Symbolically, the survival probability $W$ can be written as
\begin{equation}
\label{eq:a5}
W \; = \; S^2 T^2.
\end{equation}
$S^2$ is the probability that the gaps are not filled by secondary particles generated by 
soft rescattering, i.e.\ that no other interactions occur except the hard production process, 
see, for instance, \cite{DKS,Bjorken,FS,KMR1,EML,GLM,KMR6,KKMR,KMR7}.  The second factor, $T^2$, 
is the price to pay for not having gluon radiation in the hard production subprocess (the 
so-called bremsstrahlung fee).  It is related to classic Sudakov-suppression phenomena and is 
incorporated in the perturbative QCD calculation of the exclusive production amplitude.  The 
soft survival factor $S^2$ is the \lq\lq Achilles heel\rq\rq\ of all calculations of the 
rates of double-diffractive processes, since $S^2$ strongly depends on the phenomenological 
models for soft diffraction.  This factor is not universal, but depends on the particular hard 
subprocess, as well as on the distribution of partons inside the proton in impact parameter 
space \cite{KMR2,GLM,KMR6}.  It has a specific dependence on the characteristic momentum 
fractions carried by the active partons in the colliding hadrons \cite{KKMR}.

In \cite{KMR2} we tabulated our results for the LHC using the optimistic estimate $S^2 = 0.1$, 
whereas our detailed recent calculations \cite{KMR6} yield a lower value, $S^2 = 0.02$ 
($S^2 = 0.05$ for the Tevatron).  However it has been pointed out \cite{KMR2,KMR7} that 
it is possible to check the value of $S^2$ by observing double-diffractive dijet 
production\footnote{A promising idea of probing the gap survival factor in $Z$ production by 
$WW$-fusion, with a rapidity gap on either side of the $Z$ was advocated in Ref.~\cite{CZ}.}.  
This process is driven by the same dynamics, and has a higher cross section, so 
a comparison of the measurements with the predictions can determine $S^2$.  The recent CDF 
data for double-diffractive dijet production \cite{CDF1} appear to be consistent with our 
determination of $S^2$ in Ref.~\cite{KMR7}.  This is clear evidence in favour of the strong 
suppression due to the low survival probability of the rapidity gaps.  Further evidence 
strongly supporting the adopted description of the rapidity gap physics\footnote{Note that 
earlier CDF results \cite{Goulianos} on diffractive $W$ boson, dijet, $b$-quark and $J/\psi$ 
production rates, using forward rapidity gap tagging, have already provided evidence against 
approaches which overlook rescattering effects.} comes from a good agreement (both in 
normalisation and shape) between the CDF measurements of the diffractive dijet distribution in 
the events with a leading antiproton \cite{CDF2} and our expectations in Refs.~\cite{KKMR,KMR8}.  
Note that the observed relative suppression of Tevatron to HERA diffractive rates was 
predicted by Goulianos within the so-called renormalized Pomeron flux model \cite{KG}.

The $p\porpbar \rightarrow p + H + \porpbar$ cross section, corresponding to the basic 
mechanism shown in Fig.~1, has been calculated to single $\log$ accuracy \cite{KMR2}.  The 
amplitude is
\begin{equation}
\label{eq:a6}
{\cal M} \; = \; A \pi^3 \int \frac{d^2 Q_T}{Q_T^4} \: f_g (x_1, x_1^\prime, Q_T^2, M_H^2/4) 
\: f_g (x_2, x_2^\prime, Q_T^2, M_H^2/4),
\end{equation}
where the $gg \rightarrow H$ vertex factor $A^2$ is given by eq.~(\ref{eq:a10}) below, and 
where the unintegrated skewed gluon densities are related to the conventional distributions by 
\begin{equation}
\label{eq:a7}
f_g (x, x^\prime, Q_T^2, M_H^2/4) \; = \; R_g \: \frac{\partial}{\partial \ln Q_T^2} \: 
\left [\sqrt{T (Q_T, M_H/2)} \: xg (x, Q_T^2) \right ].
\end{equation}
The factor $R_g$ is the ratio fo the skewed $x^\prime \ll x$ integrated gluon distribution 
to the conventional one \cite{SGMR}.  $R_g \simeq 1.2(1.4)$ at LHC (Tevatron) energies.  

The bremsstrahlung survival probability $T^2$ in (\ref{eq:a5}) is given by
\begin{equation}
\label{eq:a8}
T (Q_T, \mu) \; = \; \exp \left (- \: \int_{Q_T^2}^{\mu^2} \: \frac{dk_T^2}{k_T^2} \: 
\frac{\alpha_S (k_T^2)}{2\pi} \: \int_0^{1-k_T/\mu} \: dz \: \left [ z \: P_{gg}(z) \: + \: 
\sum_q \: P_{qg}(z) \right ] \right ).
\end{equation}
The origin of the factor $1/Q_T^4$ in the integrand in (\ref{eq:a6}) reflects the fact that the 
production is mediated by the \lq fusion\rq\ of two colourless dipoles of size $d^2 \sim 1/Q_T^2$.  
Due to the presence of this factor it is argued (see e.g.\ \cite{EML}) that a perturbative 
treatment of process (\ref{eq:a1}) is inappropriate.  However, it is just the Sudakov suppression 
factor $T$ in the integrand, which makes the integration infrared stable and hence the 
perturbative predictions reliable\footnote{Moreover, the effective anomalous dimension, 
$\gamma$, of the gluon distribution $(xg (x, Q_T^2) \sim (Q_T^2)^\gamma)$ additionally 
suppresses the contribution from the low $Q_T^2$ domain \cite{KMR1}.}.  The saddle points of 
the integrand are located near $Q_T^2 = 3.2(1.5)~{\rm GeV}^2$ at LHC (Tevatron) energies.

The bremsstrahlung factor $T$ determines the probability {\it not} to emit the 
gluons in the interval $Q_T \lapproxeq k_T \lapproxeq M_H/2$.  The upper  
bound of $k_T$ is clear, and the lower bound occurs because there is destructive interference 
of the amplitude in which the bremsstrahlung gluon is emitted from a \lq\lq hard\rq\rq\ 
gluon with that in which it is emitted from the screening gluon.  That is, there is no emission 
when $\lambda \simeq 1/k_T$ is larger than the separation $d \sim 1/Q_T$ of the two $t$-channel 
gluons in the transverse plane, since then they act effectively as a colour-singlet system.

In reality, both the camels and the needle's eye, are elastic.  A camel tends to be enlarged 
by the $1/(Q_T)^4$ singularity in the integrand in (\ref{eq:a6}), while the size of the 
eye is regulated by the QCD bremsstrahlung effects in the wavelength interval $1/M_H \lapproxeq 
\lambda \lapproxeq 1/Q_T$.  The competition between these two tendencies results in the 
position of the saddle point of the integrand in (\ref{eq:a6}), which, in turn, selects the 
\lq right\rq\ camels.  Regrettably, the important dampening factor $T$ has been neglected in 
practically all theoretical papers on the double-diffractive Higgs or dijet 
production\footnote{The only exceptions are our results in Refs.~\cite{KMR1,KMR2,MRK} and the 
evaluation presented in \cite{EML}.  However there is a clear difference in the estimate of 
the survival factor $T^2$ even between these two groups.  The calculation in \cite{KMR1,KMR2,MRK} 
yields a significantly lower value of $T^2$ than that advocated in \cite{EML}.}.

The amplitude (\ref{eq:a6}) corresponds to the exclusive process (\ref{eq:a1}).  The 
modification for the inclusive process
\begin{equation}
\label{eq:a9}
pp \; \rightarrow \; X \: + \: M \: + \: Y
\end{equation}
is given in \cite{KMR1,KMR2,MRK}, where it was found that the event rate is much larger.  
However, in the inclusive case the large multiplicity of secondaries poses an additional 
problem in identifying the Higgs boson.

\section{Signal-to-background ratio for double-diffractive Higgs production}

In order to use the \lq missing-mass\rq\ method to search for an intermediate mass Higgs 
boson, via the $H \rightarrow b\bar{b}$ decay mode, we have to estimate the QCD background 
which arises from the production of a pair of jets with invariant mass about $M_H$, see 
Ref.~\cite{KMR3} for details.  Surprisingly, the critical issue of the signal-to-background 
ratio has never been addressed prior to Refs.~\cite{KMR2,KMR3} in the, otherwise, rather vast 
literature on double-diffractive Higgs production.  The good news is that the signal-to-background 
ratio does not depend on the uncertainty in the soft survival factor $S^2$, and is given just by 
the ratio of the corresponding $gg \rightarrow H \rightarrow b\bar{b}$ and $gg \rightarrow 
b\bar{b}$ subprocesses. 

We begin by assuming that the $b$ jets are not tagged.  Then the main background is the 
double-diffractive colour-singlet production of a pair of high $E_T$ gluons.  This 
background should be compared to the double-diffractive $gg \rightarrow H$ signal, with 
vertex specified by
\begin{equation}
\label{eq:a10}
\frac{A^2}{4} \; = \; \frac{\sqrt{2}}{36 \pi^2} \: G_F \: \alpha_S^2.
\end{equation}
First of all, in order to reduce the background we impose a jet $E_T$-cut.  For instance, if we 
trigger on events containing a pair of jets with angles $\theta > 60^{\circ}$ from the proton 
direction in the Higgs rest frame, then we only eliminate one half of the signal, whereas the 
background dijet cross section is \cite{KMR3}
\begin{equation}
\label{eq:a11}
\frac{d \hat{\sigma}}{dM^2} \; = \; 9.7 \: \frac{9 \alpha_S^2}{8 M^4}.
\end{equation}
Here $M$ is the invariant mass of the dijet system.

With a common scale for the coupling $\alpha_S$, and neglecting the NLO corrections, we obtain 
a signal-to-background ratio
\begin{equation}
\label{eq:a12}
\frac{S}{B_{gg}} \; = \; (4.3 \times 10^{-3}) \: {\rm Br} (H \rightarrow b\bar{b}) \: 
\left ( \frac{M}{100~{\rm GeV}} \right )^3 \: \left ( \frac{250~{\rm MeV}}{\Delta M} \right ).
\end{equation}
If $M_H = 120$~GeV, the ratio $S/B_{gg} \sim 5 \times 10^{-3}$.  This is too small for the above 
approach to provide a viable signal for the detection of the Higgs boson.  However the situation 
is greatly improved if we are able to identify $b$ and $\bar{b}$ jets.  If we assume that there 
is only a 1\% chance to misidentify a gluon jet as a $b$ jet, then tagging {\it both} the $b$ 
and $\bar{b}$ jets will suppress the gluon background by $10^4$.  In this case only the true 
$b\bar{b}$ background may pose a problem.

A remarkable advantage of the double-rapidity gap signature for the $H \rightarrow bb$ events 
is that here the $H \rightarrow b\bar{b}$ signal/$b\bar{b}$ background ratio is strongly 
enhanced due to colour factors, gluon polarization selection and the spin $\frac{1}{2}$ 
nature of quarks \cite{KMR3}.  First, the background $b\bar{b}$-dijet rate is suppressed 
due to the absence of the colour-octet $b\bar{b}$-state.  Thus, for $E_T^2 < M^2/4$ we have
\begin{equation}
\label{eq:a13}
\frac{d \hat{\sigma} (gg \rightarrow b\bar{b})}{d\hat{\sigma} (gg \rightarrow gg)} \; < \; 
\frac{1}{4 \times 27} \; < \; 10^{-2}.
\end{equation}
Second, we emphasize that for the exclusive process the initial $gg$ state obeys special 
selection rules.  Besides being a colour-singlet, for forward outgoing protons, there is a strong 
correlation between the polarizations of two incoming gluons.  Namely the fusion occurs only from 
the state with the projection of the total angular momentum $J_z = 0$ along the beam axis.  On 
the other hand, the Born amplitude for light fermion pair production\footnote{For light quark 
pair exclusive production $p + p \rightarrow p + q\bar{q} + p$, with forward outgoing protons, 
the cancellation was first observed by Pumplin \cite{Pumplin}, see also \cite{MRK,BC}.} vanishes 
in this $J_z = 0$ state, see, for example, \cite{INOK}.  This result follows from $P$- and 
$T$-invariance and fermion helicity conservation of the $J_z = 0$ amplitude \cite{DKSO}.  Thus, 
if we were to neglect the $b$-quark mass $m_b$, then at leading order we would have no QCD 
$b\bar{b}$-dijet background at all.

Of course, a non-vanishing background is expected when we allow for non-forward $b\bar{b}$ 
production due to the $|J_z| = 2$ admixture.  However such effects appear to be very small 
\cite{KMR3}.  A $b\bar{b}$ background can be also caused by the quark mass or if we emit 
an extra gluon.  Nevertheless in the former case we still have an additional suppression to 
(\ref{eq:a13}) of about a factor of $m_b^2/p_T^2 \simeq 4m_b^2/M_H^2 < 10^{-2}$, whereas in 
the latter case the extra suppression is about $\alpha_S/\pi \simeq 0.05$.  Note that events 
containing the third (gluon) jet may be experimentally separated from Higgs decay, where the 
two jets are dominantly co-planar\footnote{The situation here is similar to the signal-to-background 
ratio for intermediate mass Higgs production in polarised $\gamma\gamma$ collisions, which was 
studied in detail in \cite{DKSO,MSK}.}.  However, the price to pay for this separation is the 
further reduction of the signal caused by the additional Sudakov suppression of final state 
radiation in Higgs events \cite{DKSO}.

Thus, the two-gluon fusion mechanism for hard production, illustrated in Fig.~1, provides a 
unique situation where the polarizations of the incoming two gluons are strongly correlated 
and only helicity zero transition occurs.  An explicit calculation \cite{KMR3}, assuming $M_H 
= 120$~GeV and imposing the $\theta > 60^\circ$ cut of low $E_T$ jets, gives a signal-to-background 
ratio
\begin{equation}
\label{eq:a14}
\frac{S}{B_{b\bar{b}}} \; \gapproxeq \; 4 \: \left ( \frac{1~{\rm GeV}}{\Delta M} \right ).
\end{equation}
The signal is, thus, in excess of background even at mass resolution $\Delta M \sim 2$~GeV, so the 
$b\bar{b}$ background should not be a problem.  Unfortunately the situation worsens for 
inclusive Higgs production, where the polarization arguments become redundant.  In this case 
$S/B_{b\bar{b}}$ ratio is additionally suppressed by a factor $\sim 20$--30.

Here I must confess that just recently the $J_z = 0$ selection rule has deceived us.  
When considering in Ref.~\cite{KMR3} the double-diffractive $P$-wave heavy 
quarkonium production we overlooked the important point that in the non-relativistic limit 
the helicity-zero amplitude of the gluon-gluon coupling for the $2^{++}$ state vanishes, see, 
for example, \cite{KK}.  Actually, such suppression was first discovered in the fifties, 
in the QED context for the $2^{++}$ positronium within the non-relativistic approach of 
Ref.~\cite{Alexseev}.

After accounting for relativistic effects the $J_z = 0$ amplitude for the $2^{++} \rightarrow 
2g/2\gamma$ transition no longer vanishes, although it remains relatively small \cite{BHS}.  
Therefore the results for double-diffractive exclusive tensor $\chi$-meson production given in 
\cite{KMR3} are strongly overestimated\footnote{Vanishing of the forward double-diffractive 
$\chi (2^{++})$ production process has been recently pointed out by F.\ Yuan \cite{Yuan}, who used 
the non-relativistic formulas for derivation of $P$-wave quarkonium production amplitudes.}.

In practice, exclusive tensor $\chi$-meson production will occur due to supposedly small 
corrections caused by relativistic effects in $2^{++}$ quarkonium, as well as by the off-mass-shell 
corrections to the $J_z = 0$ transition and by the admixture of $|J_z| = 2$ di-gluon states 
induced by the non-forward contributions.  Unfortunately these later effects are less infrared 
safe than the leading pieces \cite{KMR3} and are less perturbatively controllable.  Let us 
recall that the helicity-zero suppression becomes redundant for inclusive quarkonium 
production.  It is worthwhile to mention that the axial vector ($1^{++}$) heavy quarkonium 
double-diffractive process is strongly suppressed due to Landau-Yang theorem \cite{Landau}.

\section{Rates for rapidity gap Higgs production}

While the predictions for the $S/B_{b\bar{b}}$-ratio look quite favourable for Higgs searches 
using the missing-mass method, the expected event rate casts a shadow on the feasibility of this 
approach (at least for experiments at the Tevatron).  The cross section for exclusive 
double-diffractive Higgs production at Tevatron and LHC energies has been calculated by 
several authors\footnote{A more complete set of references to related theoretical papers can 
be found in Ref.~\cite{AR}.} \cite{BL,CH,KMR1,KMR2,EML,KL}.  In our recent analysis \cite{KMR6} 
the gap survival probability for the double-diffractive process is estimated to be $S^2 = 0.05$ 
at $\sqrt{s} = 2$~TeV and $S^2 = 0.02$ at $\sqrt{s} = 14$~TeV.  If we incorporate these estimates 
into the perturbative QCD calculations \cite{KMR2}, we find
\begin{eqnarray}
\label{eq:a15}
\sigma_H \; = \; \sigma (p\bar{p} \rightarrow p + H + \bar{p}) \; \simeq \; 0.06~{\rm fb} \quad & 
{\rm at} & \quad \sqrt{s} \; = \; 2~{\rm TeV}, \\
\label{eq:a16}
\sigma_H \; = \; \sigma (pp \rightarrow p + H + p) \; \simeq \; 2.2~{\rm fb} \quad & {\rm at} & 
\quad \sqrt{s} \; = \; 14~{\rm TeV}
\end{eqnarray}
for a Higgs boson of mass 120~GeV.  These values correspond to the cross section ratio
\begin{equation}
\label{eq:a17}
\sigma_H/\sigma_{{\rm incl} (gg \rightarrow H)} \; \simeq \; 10^{-4}
\end{equation}
(c.f.\ eqs.~(\ref{eq:a2},\ref{eq:a3})), and are much lower than the predictions of other authors 
listed in \cite{AR}.

However, as we already mentioned, the recent CDF study of diffractive dijet production 
\cite{CDF1}, provides strong experimental evidence in favour of our pessimistic estimates of the 
survival factor $S^2$, see \cite{KMR7,KMR8}.  CDF \cite{CDF1} have studied double-diffractive 
dijet production for jets with $E_T > 7$~GeV.  They find an upper limit for the cross section of  
$\sigma$(dijet) $< 3.7$~nb, as compared to our prediction of about 1~nb \cite{KMR7}.  Using 
the dijet process as a monitor thus rules out the much larger predictions for $\sigma (p\bar{p} 
\rightarrow p + H + \bar{p})$ which exist in the literature.  Unfortunately the prediction 
$\sigma_H \simeq 0.06$~fb of (\ref{eq:a15}) means that Run II of the Tevatron, with an 
integrated luminosity of ${\cal L} = 15~{\rm fb}^{-1}$, should yield less than an event.

We emphasize that such a low expected signal cross section at the Tevatron just illustrates 
the high price to be paid for improving the $S/B_{b\bar{b}}$ ratio by selecting events with 
double rapidity gaps.  On the other hand, a specific prediction of the perturbative approach is 
that the cross section $\sigma_H$ steeply grows with energy \cite{KMR1,KMR2} (c.f.\ (\ref{eq:a15}) 
with (\ref{eq:a16})), in contrast to non-perturbative phenomenological models based on Ref.~\cite{BL}.  
In fact, if we were to ignore the rapidity gap survival probability, $S^2$, then $\sigma_H$ would 
have increased by more than a factor of 100 in going from $\sqrt{s} = 2$~TeV to $\sqrt{s} = 14$~TeV.  
However, at the larger energy, the probability to produce secondaries which populate the gaps 
increases, and as a result the $\sigma (pp \rightarrow p + H + p)$ increases only by a factor of 40.  
Nevertheless, there is a real chance to observe double-diffractive Higgs production at the LHC, 
since both the cross section and the luminosity are much larger than at the Tevatron.  For an 
integrated luminosity of 100~fb$^{-1}$ at the LHC we expect, for $M_H = 120$~GeV, about 200 $H 
\rightarrow b\bar{b}$ events with a favourable signal-to-background ratio.  Another test of our 
perturbative scenario is the behaviour of the dijet cross section with the jet $E_T$.  Due to 
the $x$ dependence of the perturbative gluon, we predict a steeper fall off with increasing $E_T$ 
than the non-perturbative models \cite{KMR2}.

For the inclusive production of a Higgs of mass $M_H = 120$~GeV we expect, at the LHC energy, a 
cross section of the order of 40(4)~fb, taking rapidity gaps $\Delta \eta = 2(3)$ \cite{KMR2}.

The double-diffractive dijet cross sections are much larger than those for Higgs production.  
For example, if we take a dijet bin of size $\delta E_T = 10$~GeV for each jet and $\eta_1 = 
\eta_2$ we obtain, for $E_T = 50$~GeV jets at LHC energies,
\begin{equation}
\label{eq:a18}
\left . d\sigma_{\rm excl}/d\eta \right |_0 \; \simeq \; 40~{\rm pb}, \quad\quad 
\left . d\sigma_{\rm incl}/d\eta \right |_0 \; \simeq \; 250~{\rm pb},
\end{equation}
where $\eta \equiv (\eta_1 + \eta_2)/2$.  The rapidity gaps are taken to be $\Delta \eta$(veto) $= 
(\eta_{\rm min}, \eta_{\rm max}) = (2,4)$ for the inclusive case (see \cite{MRK} for the 
definition of the kinematics).

Such a high event rate and the remarkable purity of the di-gluon system, that is generated in 
the exclusive double-diffractive production process, provides a unique environment to make 
a detailed examination of high energy gluon jets.  Note that in exclusive high-$E_T$ dijet events 
the jets appear to be pure gluon ones at the level about 3000:1.  Moreover, after an appropriate 
selection of the two-jet configuration and the removal of the $b\bar{b}$ contamination by tagging, 
the sample may become (at least) an order of magnitude purer.  Indeed, we may speak here of a 
\lq gluon factory\rq\ \cite{KMR3}.

Let us make the final comment on the soft suppression factor $S^2$, which has caused the main 
uncertainty in various calculations of the rate of rapidity gap events in the exclusive and 
inclusive processes of (\ref{eq:a1}) and (\ref{eq:a9}) respectively.  This factor depends 
sensitively on the spatial distributions of partons inside the proton, and thus, is closely 
related to the whole diffractive part of the $S$-matrix, see Ref.~\cite{KKMR} for details.  
For example, the survival probability for central Higgs production by $WW$ fusion, with large 
rapidity gaps on either side, is expected to be larger than that for the double-Pomeron 
exchange \cite{KMR2,KKMR}.  Thus, within the approach of Ref.~\cite{KKMR} the soft suppression 
factor is predicted to be $S^2 \simeq 0.25$ for the $WW \rightarrow H$ fusion at the LHC, 
which is about a factor of 10 above the estimate for the $PP \rightarrow H$ case.

Recall that a quantitative probe of the suppression factor $S^2$ can be achieved either by 
measurements of central $Z$ production \cite{CZ} or of dijet production \cite{KMR2}.

An instructive example is Higgs boson production by the $\gamma\gamma \rightarrow H$ fusion subprocess.  
This process takes place at very large impact parameters, where the corresponding gap survival 
probabilities are $S^2 = 1$ \cite{KMR2,KMOR}.  $\sigma (\gamma\gamma \rightarrow H)$ is estimated to 
be about 0.03~fb at $\sqrt{s} = 2$~TeV and 0.3~fb at $\sqrt{s} = 14$~TeV, which is comparable to our 
expectations (\ref{eq:a15}, \ref{eq:a16}) for double-diffractive Higgs production.  Note that the 
strong and electromagnetic contributions have negligible interference, because they occur at quite 
different values of the impact parameter.

\section{Conclusions}

We have examined the possibility of performing a high resolution missing-mass search for the Higgs 
boson at the Tevatron, that is the process $p\bar{p} \rightarrow p + H + \bar{p}$ where a \lq plus\rq\ 
denotes a large rapidity gap.  We find that there is a huge QCD background arising from 
double-diffractive dijet production.  A central detector to trigger on large $E_T$ jets is essential.  
Even so, the signal-to-background ratio is too small for a viable \lq missing-mass\rq\ Higgs search.  
The situation is much improved if we identify the $b$ and $\bar{b}$ jets.  The $gg \rightarrow H 
\rightarrow b\bar{b}$ signal is now in excess of the QCD $gg \rightarrow b\bar{b}$ background, even 
for a mass resolution of $\Delta M \sim 2$~GeV.  The only problem is that the survival probability of 
the gaps takes its toll, and that as a consequence the $p\bar{p} \rightarrow p + H + \bar{p}$ event 
rate appears to be too small at the Tevatron.  Note that remarkable agreement between the CDF diffractive 
dijet data \cite{CDF1,CDF2} and our predictions \cite{KKMR}-\cite{KMR8} can be viewed as a 
strong confirmation of the validity of our results for the gap survival probabilities.  In 
particular, the normalization discrepancy between the CDF diffractive results on $W$, dijet, $b$-quark 
and $J/\psi$ production \cite{CDF2,Goulianos} and the expectations based on factorization, which is 
$O$ (0.1), clearly favours low values of $S^2$.  Nevertheless, there is a real chance to observe 
double-diffractive Higgs production at the LHC, since both the cross section and the luminosity are 
much larger than at the Tevatron.

The rather pessimistic expectations of the missing-mass Higgs search at the Tevatron are, however, 
compensated by a by-product of the double-diffractive proposal.  The double-diffractive production 
of dijets offers a unique {\it gluon factory}, generating huge numbers of essentially pure gluon 
jets from a colour-singlet state in an exceptionally clean environment \cite{KMR3}.

\section*{Acknowledgements}

It is a pleasure for me to thank Giorgio Bellettini, Giorgio Chiarelli and Mario Greco for their 
kind invitation and hospitality in La Thuile.  I wish to thank Aliosha Kaidalov, Alan Martin and 
Misha Ryskin for fruitful collaboration and helpful discussions.  I am grateful to The Leverhulme 
Trust for a Fellowship.  This work was also partly supported by the EU Framework TRM programme, 
contract FMRX-CT98-0194 (DG-12-MIHT).

\end{document}